\documentclass[fleqn,usenatbib]{mnras}

\usepackage{newtxtext,newtxmath}
\usepackage{amsmath}
\usepackage{graphics,graphicx}
\usepackage{natbib}
\usepackage{setspace}
\usepackage{subcaption}
\usepackage{caption,setspace}
\newcommand{\angstrom}{\mbox{\normalfont\AA}}

\newcommand{\cch}[1]{}

\usepackage{enumitem}
\usepackage{lipsum}
\setlist[itemize]{leftmargin=*}

\usepackage{floatrow}
\usepackage[outercaption]{sidecap}    
\sidecaptionvpos{figure}{t}

\usepackage[T1]{fontenc}
\usepackage{ae,aecompl}
\pdfoutput=1

\title[Dusty simulated galaxies in FIRE-2]{Predictions for the spatial distribution of the dust continuum emission in $\mathbf{1<z<5}$ star-forming galaxies}
\author[R.K. Cochrane et al.]{R. K. Cochrane, $^{1}$\thanks{E-mail: rcoch@roe.ac.uk}
C. C. Hayward,$^{2}$
D. Angl\'es-Alc\'azar,$^{2}$
J. Lotz,$^{3}$
T. Parsotan,$^{4}$ \newauthor
X. Ma,$^{5,6}$
D. Kere\v s, $^{7}$
R. Feldmann, $^{8}$
C. A. Faucher-Gigu\`ere $^{9}$
and P. F. Hopkins $^{6}$ \\
$^{1}$SUPA, Institute for Astronomy, Royal Observatory Edinburgh, EH9 3HJ, UK\\
$^{2}$Center for Computational Astrophysics, Flatiron Institute, 162 Fifth Avenue, New York, NY 10010, USA \\
$^{3}$Space Telescope Science Institute, 3700 San Martin Drive, Baltimore, MD 21218, USA \\
$^{4}$Department of Physics, Oregon State University, 301 Weniger Hall, Corvallis, OR 97331, U.S.A.\\
$^{5}$Department of Astronomy and Theoretical Astrophysics Center, University of California Berkeley, Berkeley, CA 94720, USA\\
$^{6}$TAPIR, MC 350-17, California Institute of Technology, Pasadena, CA 91125, USA\\
$^{7}$Department of Physics, Center for Astrophysics and Space Sciences, University of California, San Diego, La Jolla, CA, USA\\
$^{8}$Institute for Computational Science, University of Zurich, Zurich CH- 8057, Switzerland\\
$^{9}$Department of Physics and Astronomy and CIERA, Northwestern University, Evanston, IL 60208, USA}
\date{Accepted 2019 May 8. Received 2019 May 3; in original form 2019 March 18.}

\pubyear{2019}
\volume{{\rm in press}}

\begin{document}
\label{firstpage}
\pagerange{\pageref{firstpage}--\pageref{lastpage}}
\maketitle

\begin{abstract}
We present the first detailed study of the spatially-resolved dust continuum emission of simulated galaxies at $1<z<5$. 
We run the radiative transfer code {\sc skirt} on a sample of submillimeter-bright galaxies drawn from the Feedback in Realistic Environments (FIRE) project. These simulated galaxies reach Milky Way masses by $z=2$. Our modelling provides predictions for the full rest-frame far-ultraviolet-to-far-infrared spectral energy distributions of these simulated galaxies, as well as 25-pc-resolution maps of their emission across the wavelength spectrum. The derived morphologies are notably different in different wavebands, with the same galaxy often appearing clumpy and extended in the far-ultraviolet yet an ordered spiral at far-infrared wavelengths. The observed-frame 870-\micron\, half-light radii of our FIRE-2 galaxies are $\sim0.5-4\,\rm{kpc}$, consistent with existing ALMA observations of galaxies with similarly high redshifts and stellar masses. In both simulated and observed galaxies, the dust continuum emission is generally more compact than the cold gas and the dust mass, but more extended than the stellar component. The most extreme cases of compact dust emission seem to be driven by particularly compact recent star-formation, which generates steep dust temperature gradients. Our results confirm that the spatial extent of the dust continuum emission is sensitive to both the dust mass and SFR distributions.
\end{abstract}

\section{Introduction}
Observations suggest that the physical properties of star-forming (SF) galaxies at the peak of cosmic star-formation ($z\sim2$) differ greatly from those of the ordered disks and ellipticals in the local Universe. Galaxies at these redshifts display high star-formation rates, believed to be driven by large molecular gas reservoirs \citep{Tacconi2010,Tacconi2013,Tacconi2017,Papovich2016,Falgarone2017a,Jimenez-Andrade2018} that arise due to steady accretion of cold gas along filaments of the cosmic web \citep{Keres2005,Dekel2009,Faucher2011,Martin2016,Kleiner2017}. Structurally, high-redshift galaxies are less ordered than their low-redshift counterparts, with star-formation taking place within turbulent disks \citep{Genzel2008,Kassin2012,Guo2015,Tadaki2018} that often harbour massive ultraviolet (UV)-bright clumps \citep{Elmegreen2013,Guo2017b,Soto2017}. However, our measurements of the high-redshift Universe are largely reliant on data at rest-frame optical and UV wavelengths, which can be biased towards the least dust-obscured galaxies. \\
\indent  Only in the last few years have new facilities such as ALMA had the resolving power to resolve and probe the morphology of longer-wavelength emission from highly star-forming galaxies. The angular resolution of previous instruments such as SCUBA \citep{Holland1999} was low, so it was not possible to determine the structural properties of high-redshift galaxies. Source confusion has also been a  hindrance in the identification of fainter sources \citep[e.g. with Herschel;][]{Oliver2012,Scudder2016}. ALMA has the potential to be particularly fruitful in identifying high-redshift galaxies, due to the so-called `negative k-correction' at rest-frame far-infrared (FIR) fluxes (\textcolor{black}{observed} $\lambda\sim850$-\micron\ flux from galaxies with similar intrinsic spectra remains approximately constant across the redshift range $z\sim1-6$, as we trace further up the Rayleigh-Jeans tail at higher redshift). The most interesting physical insights will likely come from the combination of these new millimeter/sub-millimeter (mm/sub-mm) data with shorter-wavelength imaging. To this end, \cite{Dunlop2017} present $1.3\rm{mm}$ ALMA imaging of the Hubble UltraDeep Field (HUDF), previously mapped with the Wide Field Camera 3/IR on the Hubble Space Telescope (HST) to an unprecedented $5\sigma$ depth of $30\rm{AB\,mag}$ \citep{Bouwens2010,Oesch2010,Illingworth2013,Dunlop2013,Ellis2013}, and also with the HST Advanced Camera for Surveys \citep[ACS;][]{Beckwith2006}, over an area of $4.5\rm{arcmin}^{2}$. Combining these new ALMA data with Herschel and Spitzer 24-\micron\, photometry and fitting to a template spectral energy distribution (SED), \cite{Dunlop2017} find that $\sim85\%$ of the total star-formation at $z\sim2$ is enshrouded in dust (see also \citealt{Bourne2017,Whitaker2017}). They show that for high-mass galaxies ($M_{*} > 2\times10^{10}M_{\odot}$), which host $\sim65\%$ of the total star-formation at this epoch, the star-formation rate derived from long-wavelength emission is an extraordinary $200$ times that derived from unobscured light. \cite{Bowler2018} demonstrate that dust-obscured star-formation could be substantial even as early as $z\sim7$.\\
\indent In addition to finding substantially different measurements of star-formation rates compared to dust-uncorrected short-wavelength light, studies of galaxies at longer wavelengths also present a different view of the morphologies of high-redshift galaxies. Unobscured emission (probed at short wavelengths) tends to be significantly more extended and clumpier than the rest-frame FIR emission. \cite{Barro2016}, for example, find that the 345-GHz (ALMA Band 7) dust continuum emission of a $M_{*}=10^{10.9}M_{\odot},\,\,\rm{SFR}=500M_{\odot}yr^{-1},\,\,z=2.45$ galaxy has a half-light radius which is half that of the rest-frame optical emission probed by HST. \cite{Hodge2016} imaged 16 $z\sim2.5$ similarly massive, highly star-forming, luminous sub-mm galaxies at $0.16''$ in the same ALMA band. Many of these galaxies display clumpy structures in HST's $H_{160}$ and $I_{814}$ bands, but their dust emission appears substantially smoother and more compact. The 870-\micron\, radii obtained are small (median $1.8\pm0.2\,\rm{kpc}$), with no convincing evidence for clumpy dust emission at the ALMA resolution probed. \\
\indent Molecular gas reservoirs have also been mapped for a handful of high-redshift, FIR-luminous galaxies. High-spatial-resolution studies show that molecular gas is compact, though it tends to be slightly more extended than the dust continuum emission. \cite{Tadaki2017} imaged two $z=2.5$ galaxies at 345 GHz, obtaining 870-\micron\, dust-emission radii of $1.2\pm0.1\,\rm{kpc}$ and $1.3\pm0.1\,\rm{kpc}$, around half the size of the CO(J=3-2) emission. Consistent results have been found by other studies \citep[e.g.][]{Strandet2017,Rivera2018,Tadaki2018}, though sample sizes remain small. \\
\indent A physical understanding of these differences in spatial extent of emission in different wavebands is critical in order to make the best use of the unmatched sensitivity and high spatial resolution of ALMA. This is currently difficult observationally, as only small samples of galaxies have been resolved at high resolution with multiple instruments, and such samples are often biased towards either the least dust-obscured systems (if selected in the UV) or the most compact, FIR-bright systems (if selected by FIR surface brightness). Interpretations are further complicated by uncertainty in what FIR/sub-mm fluxes actually probe. Frequently used relations between observed FIR/sub-mm fluxes and SFR \citep[e.g.][]{Kennicutt2012} do not fold in the shape of the full dust SED, which should reflect not only SFR but also dust mass, dust temperature, and the geometry of the source \citep[e.g.][]{Misselt_2001,Hayward2011,Hayward2012,Lanz2014,Safarzadeh2016,Kirkpatrick2017}. \\
\indent The primary aim of this paper is to understand which physical properties of high-redshift galaxies are probed by their dust continuum emission, in a spatially-resolved manner. We address this using simulations to study individual galaxies at high spatial resolution across the rest-frame UV-FIR spectrum. We select a sample of the most FIR-bright galaxies from the Feedback In Realistic Environments 2 (FIRE-2) simulations \citep{Hopkins2017}\footnote{http://fire.northwestern.edu} presented in \cite{Angles-Alcazar2017}. We perform three-dimensional continuum radiative transfer on selected galaxy snapshots from these simulations, modelling the effects of dust attenuation and re-emission to predict the spatially-resolved multi-wavelength emission of these high-redshift FIR/sub-mm-bright simulated galaxies. We then compare the spatial extent of the dust continuum emission to various intrinsic physical properties of our simulated galaxies. \\
\indent The structure of this paper is as follows. In Section \ref{sec:simulating_gals}, we discuss our method for selecting sub-mm-bright snapshots from the FIRE-2 simulations. We describe the radiative transfer modelling used to post-process these galaxy snapshots and present their simulated rest-frame UV-FIR spectral energy distributions in Section \ref{sec:skirt_modelling}. We present predictions for the spatial extent of dust continuum emission in Section \ref{sec:dust_predictions} and also compare our predictions to observational results. We also study which physical properties of high-redshift, dusty star-forming galaxies are best probed by the spatial extent of the dust continuum emission. We summarise our conclusions in Section \ref{sec:conclusions}.

\begin{table*}
\begin{center}
\begin{tabular}{c|c|c|c|c|c}
Name & $\log_{10}M_{\rm{halo}}/M_{\odot}$ & $\log_{10}M_{\rm{*}}/M_{\odot}$ & $\log_{10}M_{\rm{gas}}/M_{\odot}$ & $\rm{SFR}/M_{\odot}\rm{yr}^{-1}$ & $r_{1/2}/\rm{kpc}$ \\ 
\hline
A1 & $12.45$ & $11.24$ & $10.22$ & 65 & 0.73 \\
A2 & $12.56$ & $11.46$ & $10.51$ & 168 & 0.98 \\
A4 & $12.49$ & $11.10$ & $10.33$ & 63 & 0.81 \\
A8 & $12.41$ & $10.85$ & $10.45$ & 88 & 0.91 \\
\hline
\end{tabular}
\caption{Properties of the four simulated FIRE-2 haloes at $z=2$.  $M_{\rm{halo}}$ denotes the total mass of the central halo at $z=2$. $M_{\rm{*}}$, $M_{\rm{gas}}$, and $\rm{SFR}$ denote the stellar mass, gas mass, and star-formation rate of the halo's central galaxy at $z=2$, all calculated within $0.1R_{\rm{vir}}$. $r_{1/2}$ is the half-mass radius, calculated using the stellar mass within $0.1R_{\rm{vir}}$.}
\label{Table:FIRE_halos}
\end{center}
\end{table*}

\section{A sample of simulated high-redshift galaxies}\label{sec:simulating_gals}
\subsection{Galaxies in the FIRE-2 simulations}\label{sec:FIRE2}
The Feedback In Realistic Environments (FIRE) project \citep{Hopkins2014,Hopkins2017} is a set of state-of-the-art hydrodynamical cosmological zoom-in simulations that explore the role of stellar feedback in galaxy formation and evolution. Stellar feedback must play an important role in galaxy formation. Without it, the galactic ISM would collapse on dynamical time-scales, leading to gravitational collapse, fragmentation and accelerated star-formation. While galaxies simulated without stellar feedback thus rapidly convert all available gas into stars (e.g. \citealt{Hopkins2011}; see the review by \citealt{Somerville}), the tight locus of observed galaxies on the Kennicutt-Schmidt relation \citep{Jr1998} implies that gas consumption timescales in real galaxies are long. Furthermore, both galaxy stellar mass functions \citep[e.g.][]{Ilbert2013,Muzzin2013,Davidzon2017} and the stellar mass-halo mass relation (e.g. \citealt{Moster2010,Behroozi2013,Cochrane2017a}) imply that galaxies convert only a small fraction of the universal baryon fraction into stars. Galactic outflows are therefore needed to regulate the mass of galaxies over time \citep{Keres2009,Angles2014,Muratov2015,angles470}. These outflows are likely responsible for the observed enrichment of the circumgalactic medium (CGM) and intergalactic medium (IGM), and are also observed directly (e.g. \citealt{Weiner2009,Steidel2010}; see the review by \citealt{Rupke2018}). These pieces of observational evidence demand that stellar feedback must be at play \citep{Schaye2003,Oppenheimer2006,Faucher2015,Hafen2017}. Stellar feedback is believed to be particularly important in low-stellar-mass galaxies, below the peak of the stellar mass-halo mass relation. \\
\indent Various stellar feedback processes are thought to contribute. These processes include supernovae, protostellar jets, photo-heating, stellar mass loss from O- and AGB-stars and radiation pressure \citep[see][for a review]{Dale2015}. Importantly, these processes are believed to act non-linearly, and so modelling the stellar processes of even a single galaxy becomes a complex computational challenge. Only recently have cosmological zoom simulations achieved sufficient resolution to model these feedback processes directly. \\
\indent The FIRE project employs two main techniques to explicitly model multi-channel stellar feedback. Firstly, the FIRE simulations resolve the formation of giant molecular clouds (GMCs), and star-formation takes place only in self-gravitating (according to the \citealt{Hopkins2013sf_criteria} criterion), self-shielding molecular gas (following \citealt{Krumholz2011}) at high densities ($n_{H}>1000\,\rm{cm}^{-3}$ in the simulations we use here).
Secondly, FIRE includes models for both energy and momentum return from the main stellar feedback processes, using directly the predictions of stellar population synthesis models without the extensive parameter tuning employed in other simulations. Specifically, once a star particle forms, the simulations explicitly follow several different stellar feedback mechanisms as detailed in \citealt{Hopkins2018feedback}, including (1) local and long-range momentum flux from radiation pressure (in both the initial UV/optical single-scattering regime and re-radiated light in the IR); (2) energy, momentum, mass and metal injection from SNe (Types Ia and II) and stellar mass loss (both OB and AGB); and (3) photo-ionization and photo-electric heating. Every star particle is treated as a single stellar population with known mass, age, and metallicity, and then all feedback event rates, luminosities and energies, mass-loss rates, and all other quantities are tabulated directly from stellar evolution models ({\sc starburst99}; \citealt{Leitherer1999}), assuming a \citet{Kroupa2001} initial mass function. The FIRE simulations succeed in generating galactic winds self-consistently \citep{Muratov2015,angles470}, 
\textcolor{black}{without relying on sub-grid hydrodynamic decoupling or delayed cooling techniques}, and broadly reproducing many observed galaxy properties, including stellar masses, star-formation histories and the `main sequence' of star-forming galaxies (see \citealt{Hopkins2014,Sparre2017}), metallicities and abundance ratios \citep{Ma2016, VandeVoort2015}, as well as morphologies and kinematics of both thin and thick disks \citep{Ma2017}. \\
\indent For this paper, we study the central galaxies of four massive haloes originally selected and simulated by \citet{Feldmann2016, Feldmann2017a} with the original FIRE model \citep{Hopkins2014} as part of the {\sc MassiveFIRE} suite. The haloes were selected to have dark matter halo masses of $M_{\rm{halo}}\sim10^{12.5} M_{\odot}$ at $z=2$. The central galaxies of these haloes have stellar masses of $7\times10^{10}-3\times10^{11}M_{\odot}$ at $z=2$ (detailed in Table \ref{Table:FIRE_halos}), with a variety of formation histories; see \cite{Feldmann2017a} for details. The simulations in the present study are drawn from \cite{Angles-Alcazar2017}, who re-simulated these four massive haloes down to $z=1$ with the upgraded FIRE-2 physics model \citep{Hopkins2017} and with a novel on-the-fly treatment for the seeding and growth of supermassive black holes (SMBHs). Note that these simulations do not include feedback from the accreting SMBHs. Compared to FIRE, FIRE-2 simulations are run with a new, more accurate hydrodynamics solver (a mesh-free Godunov solver implemented in the {\sc gizmo}\footnote{\url{http://www.tapir.caltech.edu/~phopkins/Site/GIZMO.html}} code; \citealt{Gaburov2011,Hopkins2015}). They also feature improved treatments of cooling and recombination rates, gravitational softening and numerical feedback coupling, and they adopt a higher density threshold for star-formation \citep{Hopkins2018feedback}. The mass resolution of our simulations is $3.3\times{}10^4$ $M_\odot$ for gas and star particles and $1.7\times{}10^5$ $M_\odot$ for dark matter particles.

\begin{figure*} 
	\centering
    \includegraphics[scale=0.5]{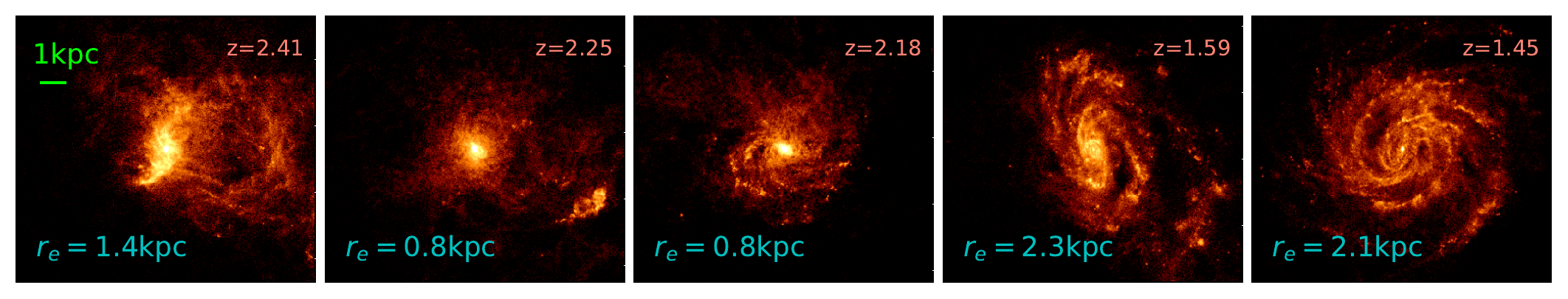}
	\caption{870-\micron~observed-frame flux maps for a subsample of snapshots of the central galaxy of FIRE-2 halo A4 predicted using the {\sc skirt} radiative transfer code \textcolor{black}{and plotted on the same logarithmic flux scale}. These cutouts span $10\rm{kpc}\times10\rm{kpc}$ (proper distances). The predicted dust continuum emission displays a range of morphologies as the ordered disk develops. This emission is compact, spanning half-light radii of $\sim1-2\,\rm{kpc}$ (these are shown in blue).}
  \label{fig:850um_images}
\end{figure*}
\subsection{Selection of sub-mm-bright galaxy snapshots at $z>2$}\label{sec:gal_props}
We wish to simulate galaxies that are representative of those typically observed with ALMA at high spatial resolution at high redshifts, which implies that we should select those that are likely to have high 850-\micron\, flux densities ($f_{850} \ga 1\,{\rm mJy}$). Performing radiative transfer on each of the $\sim300$ redshift snapshots available to predict sub-mm fluxes and then selecting the brightest would be unnecessarily computationally intensive. The first step in our analysis is therefore to select redshift snapshots for each of the four galaxies for which we expect the sub-mm flux to be particularly bright, using simply the SFR and dust mass at each redshift. We adopt the following equation, derived from fits to the \textcolor{black}{observed-frame 850-\micron} flux densities of simulated galaxies computed via dust radiative transfer and presented in \cite{Hayward2013}:
\begin{equation}\label{eq:hayward2013}
f_{850} = 0.81 \,\rm{mJy} \times \Bigg(\frac{SFR}{100M_{\odot}yr^{-1}}\Bigg)^{0.43}\Bigg(\frac{M_{\rm{dust}}}{10^{8}M_{\odot}}\Bigg)^{0.54},
\end{equation}
where we estimate the dust mass for the present purposes using $M_{\rm{dust}} = 0.01 M_{\rm{gas}}$, where $M_{\rm{gas}}$ is the total gas mass within $0.1R_{\rm{vir}}$. \textcolor{black}{The agreement between these predictions and the 850-\micron\ flux densities that result from the radiative transfer modelling is correct to within a factor of $\sim2$ for fluxes above $\sim 1\rm{mJy}$. Future work will involve running radiative transfer over a larger number of snapshots and re-fitting this formula.}
We select the $\sim20$ snapshots with the highest predicted $f_{850}$ for each simulated halo (named A1, A2, A4 \& A8)\footnote{Our analysis focuses on the central galaxies of each halo. Throughout, we will refer to the central galaxy of e.g. halo A1 simply as `galaxy A1'.}, within the redshift range $1<z<5$. A couple of snapshots were excluded from the analysis after performing radiative transfer due to poorly defined centres and extremely diffuse dust emission. We do not attempt to produce a complete sample of sub-mm bright galaxies from the FIRE-2 simulations; instead, our selection is sufficient to yield a sample of sub-mm bright snapshots for which we can perform radiative transfer and study multi-wavelength properties.

\section{Radiative transfer methods and results}\label{sec:skirt_modelling}
\subsection{Performing radiative transfer with {\sc skirt}}\label{sec:skirt_details}
Modelling dust and its emission in galaxies is a difficult computational problem \citep[see][for a comprehensive review]{Steinacker2013}. The process of radiative transfer is non-local in space (photons can propagate long distances before interacting with dust), and it is coupled in terms of both direction and wavelength. The distribution of dust in (both real and simulated) galaxies is far from a simple screen; instead, it is necessary to model the three-dimensional distribution of sources of radiation (stars and AGN) and dust.\\
\indent In this work, we make use of the Stellar Kinematics Including Radiative Transfer ({\sc skirt})\footnote{\url{http://www.skirt.ugent.be}} Monte Carlo radiative transfer code \citep{Baes2011,Camps2014}, \textcolor{black}{which has also been used to model dust attenuation and emission in the EAGLE simulations \citep[see][]{Trayford2017,Camps2018,McAlpine2019}}. Monte Carlo radiative transfer codes like {\sc skirt} treat the radiation field from stars (and sometimes AGN) as a flow of photons through the dusty medium of a galaxy to compute the effects of dust absorption, scattering, and re-emission of the absorbed light, including dust self-absorption. We are able to perform the radiative transfer in post-processing because the light-crossing time is short compared to dynamical times on resolved scales (such that the dust and stellar geometry does not change appreciably over a light-crossing time).\\
\indent We extract gas and star particles from the FIRE-2 simulations at each of our chosen snapshots. The coordinate system is rotated to align with the angular momentum vector of the gas particles within $0.1R_{\rm{vir}}$ prior to input to {\sc skirt}, so that for disk galaxies, a viewing angle of 0 degrees corresponds to face-on. For gas particles with temperature $<10^{6}\,\rm{K}$, we compute dust masses using the metallicity of the gas particles and a dust-to-metals mass ratio of $0.4$ \citep{Dwek1998,James2002}. We assume that dust is destroyed in gas particles with temperature $>10^6\,{\rm K}$ \citep{Draine1979,Tielens1994}. We use a \cite{Weingartner2001} Milky Way dust prescription to model a mixture of graphite, silicate and PAH grains. Star particles are assigned \citet{Charlot2003} SEDs according to their ages and metallicities (note that our results are unchanged if we instead use the {\sc starburst99} SED templates presented by \citealt{Leitherer1999}). We use an octree dust grid, in which cell sizes are adjusted according to the dust density distribution. We impose the condition that no dust cell may contain more than $0.0001\%$ of the total dust mass of the galaxy, which yields excellent convergence of the integrated SED. \\
 \indent We also specify a number of properties relating to the SED output. We define a wavelength grid with $\sim100$ discrete wavelengths, spaced uniformly in log(wavelength) between rest-frame UV and FIR wavelengths and including each of the ALMA bands. To model the flux that would be received by an observer on Earth, we place seven detectors at $z=0$, uniformly spaced at different inclinations with respect to the disk plane of the galaxy. These detectors have pixel sizes corresponding to a proper length of $25\,\rm{pc}$, and the field of view is set to 10\% of the virial radius for each galaxy snapshot studied. Note that the pixel scale of the images should not be confused with the resolution of the simulation; this varies across the galaxy, and is higher than $25\,\rm{pc}$ in denser regions but lower in general ISM gas. \\
 \indent The focus on resolved dust emission in the present work is complementary to other radiative transfer studies of FIRE galaxies with {\sc skirt}. \cite{Liang2018} analyzed how well galaxy-integrated, single-band dust continuum emission traces molecular gas. The implications of defining galaxy-integrated dust temperatures in different ways, for instance with regard to various scaling relations, are studied in \cite{Liang2019}. Finally, \cite{Ma2019} make predictions for the dust extinction and emission properties of $z \ga 5$ galaxies using a sample of 34 FIRE-2 haloes, including those first presented in \citet{Ma2018}.
 
\subsection{Morphology across the wavelength spectrum}\label{sec:all_wavelength_morphs}
\begin{figure} 
    \centering
    \includegraphics[scale=0.6]{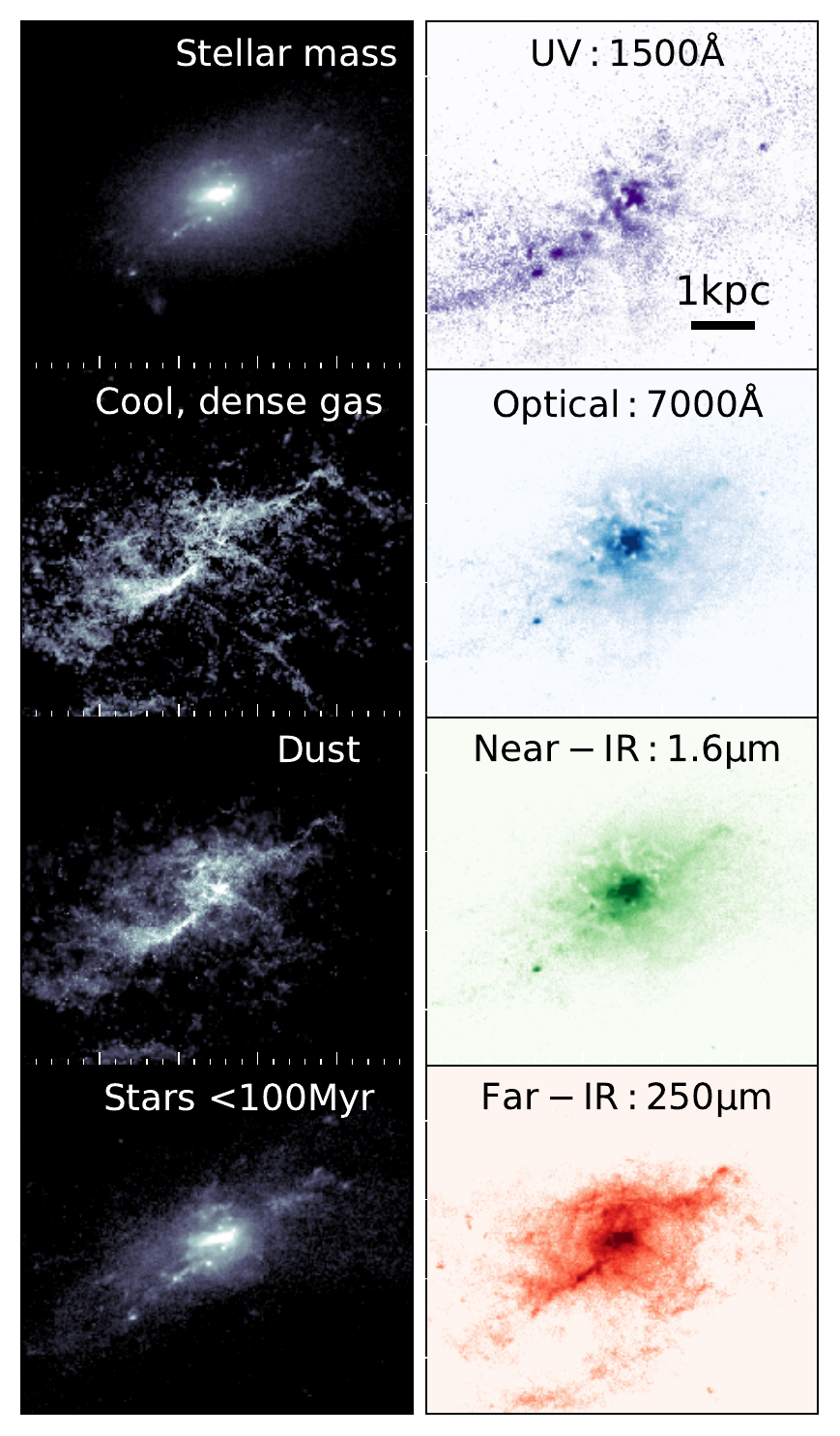}
    \caption{The wavelength-dependent morphology of galaxy A1 at $z=4.38$. The left panels show the projected distributions of stellar mass, cold, dense gas mass, dust mass, and stars formed within $100\rm{Myr}$ (intrinsic properties of the galaxy). The right panels show the {\sc skirt}-predicted images at different rest-frame wavelengths. The morphology is strongly dependent on the wavelength. The galaxy appears clumpy and extended in the rest-frame UV but more ordered at longer wavelengths. The colour scales are logarithmic and span the $70^{\rm{th}}-99^{\rm{th}}$ percentiles of the flux distribution of each panel, to highlight the qualitative differences in morphology.}
    \label{fig:morphologies_A1}
\end{figure}
\begin{figure} 
    \centering
    \includegraphics[scale=0.6]{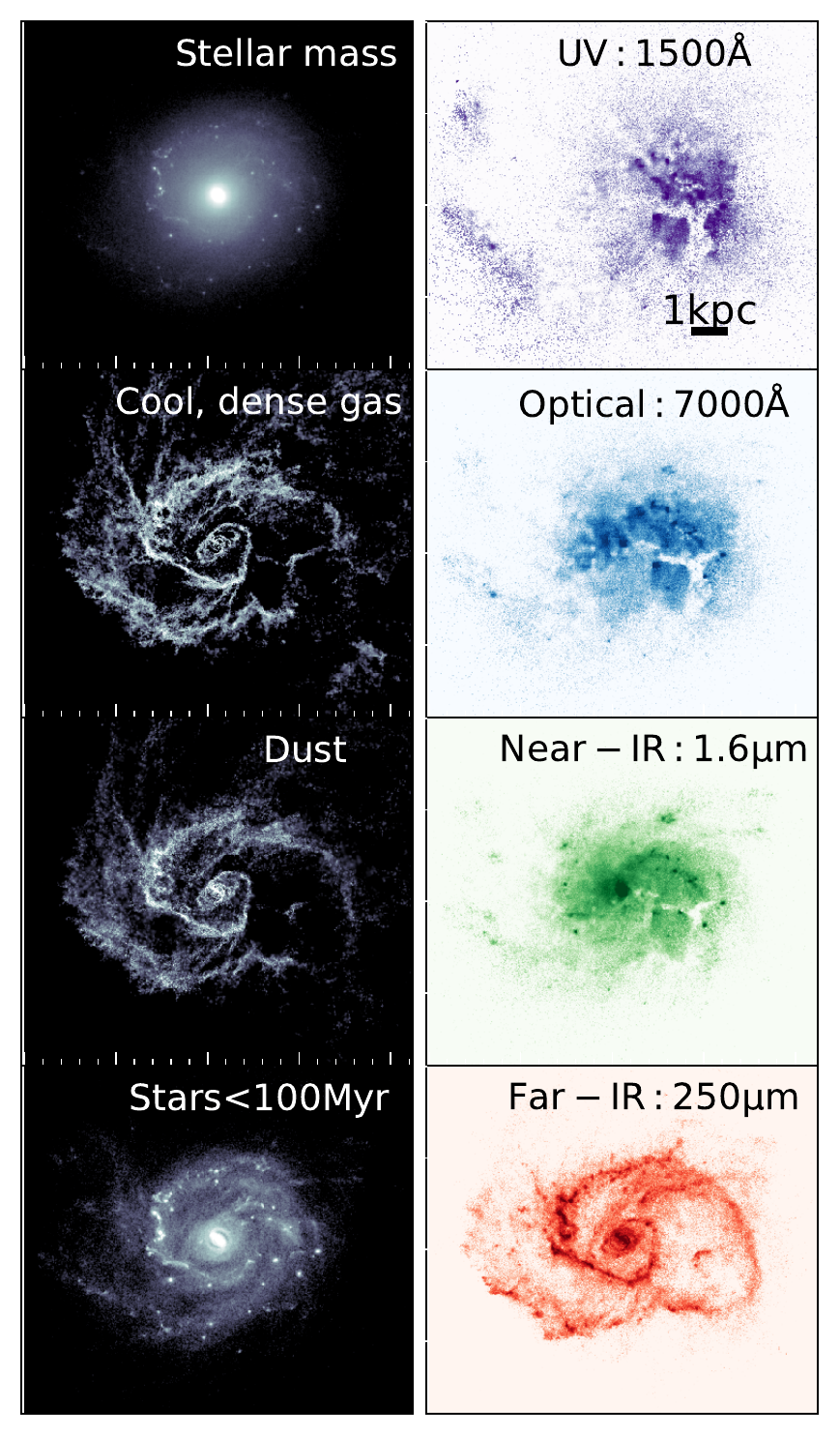}
   \caption{The wavelength-dependent morphology of galaxy A2 at $z=2.95$, with panels as described in Figure \ref{fig:morphologies_A1}. The UV and optical light is significantly offset from the peak of the stellar mass and SFR, appearing to trace holes in the dust.}
     \label{fig:morphologies_A2}
\end{figure}
\begin{figure} 
	\centering
    \includegraphics[scale=0.55]{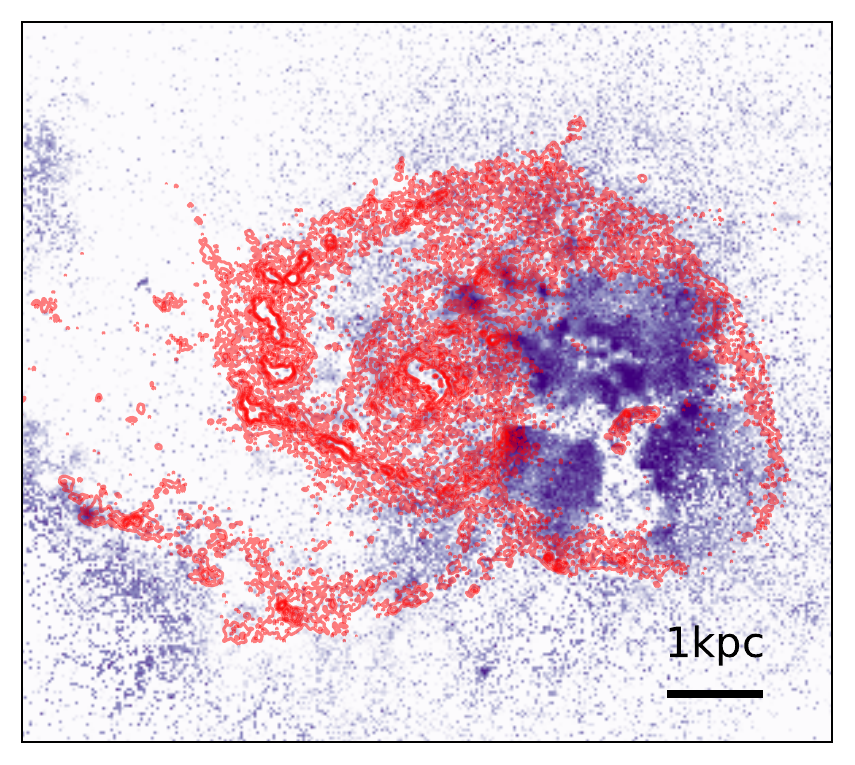}
   \caption{The rest-frame $1500\,\angstrom$ image of galaxy A2 at $z=2.95$, with 250-\micron\, contours overlaid in red. The long- and short-wavelength emission occupy strikingly different spatial regions. While there is recent star-formation across the extent of the galaxy disk, light at short wavelengths does not escape from regions of high dust density. This leads to a spatial offset between the FUV and dust continuum emission.}
     \label{fig:overlay_A2}
\end{figure}

Our radiative transfer modelling enables us to track the emission from each of the galaxies at multiple epochs. We compute images of each of the galaxy snapshots at every wavelength on our grid, spanning the rest-frame far-UV to far-IR. We show an example of the high quality of our maps of the sub-mm flux in Figure \ref{fig:850um_images}, to illustrate the wide range of morphologies displayed by a single galaxy evolving in the redshift range $z=2.41-1.45$. We also find that the same galaxy can look vastly different in the different wavebands. We illustrate this qualitatively with five representative wavelengths for the central galaxy of halo A1 (hereafter `galaxy A1') at $z=4.38$ in Figure \ref{fig:morphologies_A1} and galaxy A2 at $z=2.95$ in Figure \ref{fig:morphologies_A2}. We also show for comparison the spatial distributions of four key galaxy properties: total stellar mass, cool, dense gas mass ($T<300\,{\rm K}, n_{H} > 10\, \rm{particles}/\rm{cm}^{-3}$), dust mass, and recently formed stars ($\rm{age}<100\,\rm{Myr}$). \\
\indent Galaxy A1 at $z=4.38$ has a {\sc skirt}-predicted observed-frame 850-\micron\, flux density of $0.79\,\rm{mJy}$. At this snapshot, the galaxy exhibits very clumpy FUV emission (rest-frame $1500\angstrom$), spanning a few kpc. This emission is aligned with the cool gas and exhibits peaks where this gas is densest. The optical emission (rest-frame $4000\angstrom$) is also clumpy. Longwards of 24-\micron\, the emission becomes more regular and centrally concentrated, resembling the total and recently formed stellar mass distributions more closely. \\
\indent Galaxy A2 at $z=2.95$ is brighter, with a {\sc skirt}-predicted observed-frame 850-\micron\, flux density of $1.45\,\rm{mJy}$. It is also substantially more extended in all wavebands. The rest-frame $1500\angstrom$ and $7000\angstrom$ emission is again clumpy, bearing little resemblance to the ordered bulge plus spiral structure that is clear from the SFR and stellar, gas and dust mass maps. Interestingly, the peak of the short-wavelength emission occurs in a region to the right of centre that is largely free of dust (see Figure \ref{fig:overlay_A2}). It appears that the clumpiness of the FUV emission is driven by the structure of the dust, with FUV emission tracing holes in the dust. This is just one example, and such offsets are common in the sample, as in observations \citep[e.g.][Cochrane et al. in prep]{Chen2017,Rivera2018}. Future work will explore these offsets, as well as the resolved multi-wavelength emission on a pixel-by-pixel basis.

 \begin{figure*} 
	\centering
	\includegraphics[scale=0.54]{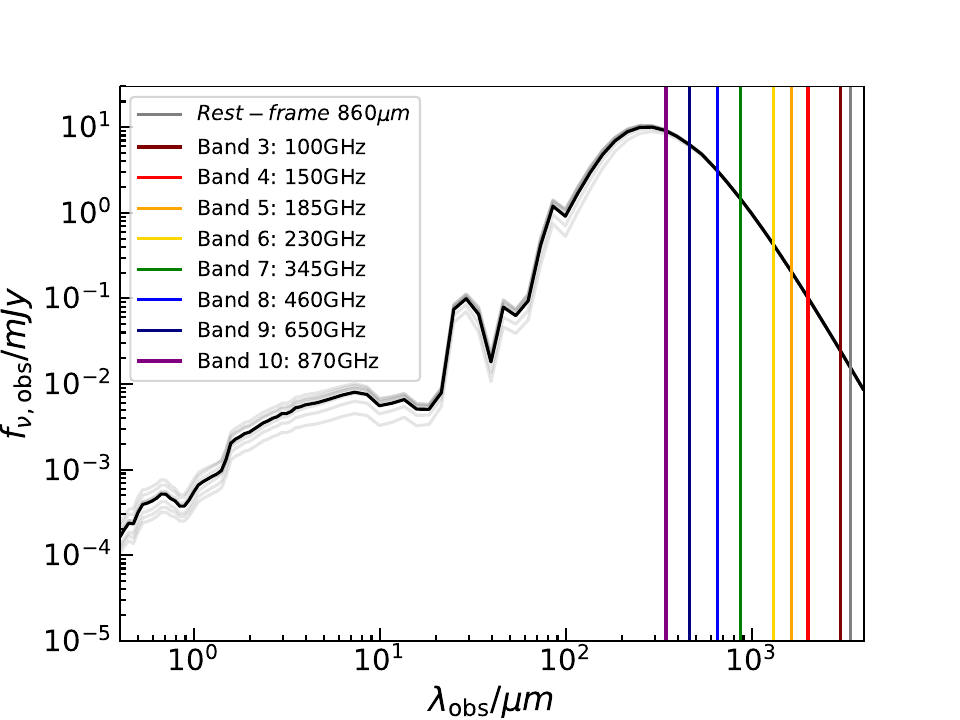}
	\includegraphics[scale=0.54]{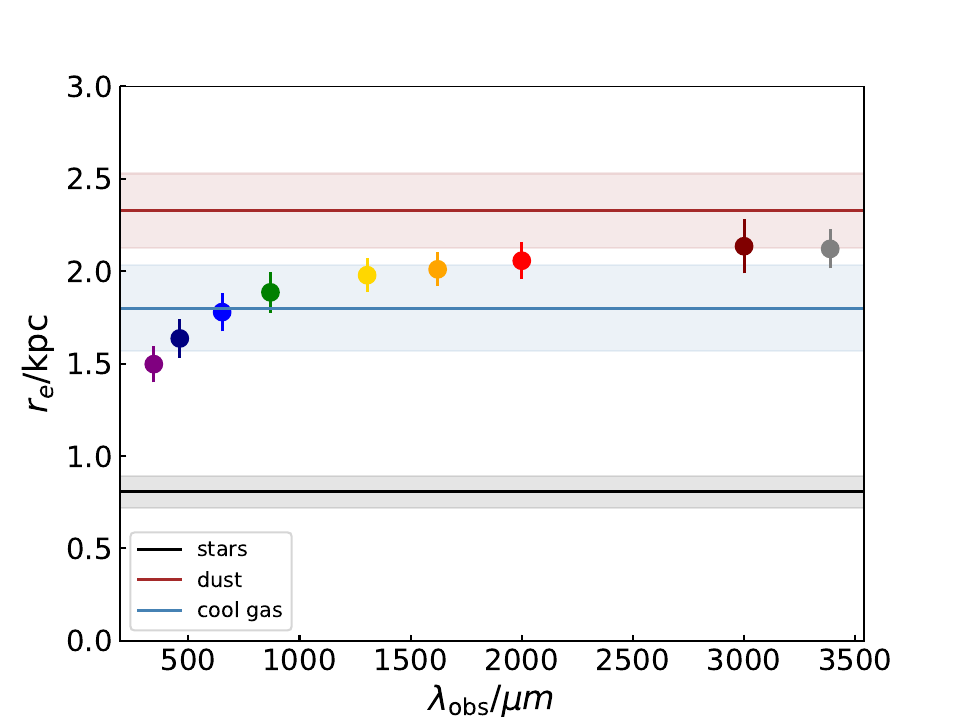}
	\caption{Left: observed-frame SED of galaxy A2 at $z=2.95$, with the centres of ALMA bands overlaid. The grey lines show the SED at different inclinations, with the mean shown in black. Right: the half-light radii of the emission at the wavelengths of centres of each of these bands compared to those of the galaxy stellar mass, cold gas mass and dust mass. Error bars and shaded regions are derived considering seven different viewing angles. Shorter-wavelength FIR emission is closer in size to the stellar component, whereas longer-wavelength emission traces the cold gas and dust.}
    \label{fig:A1_79_SED}
\end{figure*}

\begin{figure}
	\centering
	\includegraphics[scale=0.99]{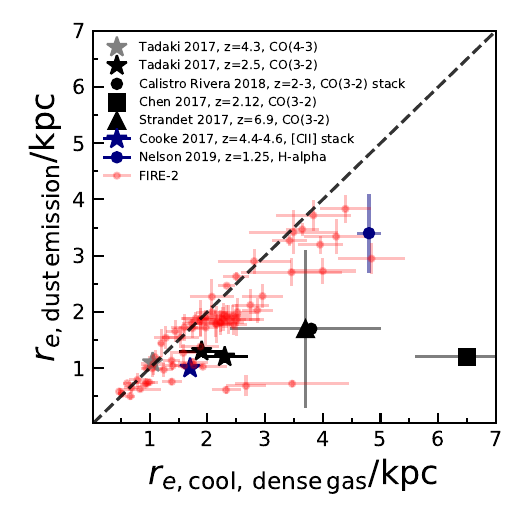}
	\vspace{-0.5cm}
	\caption{The effective radius of the 870-\micron\, dust emission versus that of the cool, dense gas ($T<300\,{\rm K}, n_{H} > 10\, \rm{particles}/\rm{cm}^{-3}$), for FIRE-2 galaxy snapshots (red) and selected observational measurements. The dashed black line shows the 1-1 relation. For all FIRE-2 galaxies, the dust emission is more compact than the cool, dense gas at almost all redshift snapshots studied, consistent with the observational sample (see text). The effective sizes of dust emission and cool gas of our simulated galaxies agree well with those of the observed galaxies, although the slope of the relation between the two sizes seems to be somewhat steeper for the simulations than for the (small) sample of observed galaxies. However, the slope of the relation for simulated galaxies is sensitive to the temperature and density cuts made to select cool, dense gas.}
    \label{fig:cool_gas_dust_observations}
\end{figure}

\section{Quantifying the spatial extent of dust continuum emission}\label{sec:dust_predictions}
\subsection{The dependence of size on FIR wavelength}
In this section, we quantify the sizes of sub-mm-bright FIRE-2 galaxies as a function of wavelength. An example of an observed-frame SED predicted by {\sc skirt}, overplotted with the wavelengths corresponding to eight ALMA bands, is shown in Figure \ref{fig:A1_79_SED} (left-hand panel). We extract the {\sc skirt} image at each of these wavelengths and derive an `effective radius' for the predicted emission. This is the radius that contains half of the total flux (calculated within a $\sim12\rm{kpc}$ radius), derived using a circular aperture centred on the flux-weighted centre of the emission in each band. In each case, an error bar is derived from the standard deviation of the effective radius measurements at seven different galaxy/detector inclinations. \\
\indent In the right-hand panel of Figure \ref{fig:A1_79_SED}, we show an example of our results. The effective radius of the emission varies with wavelength, with longer wavelength FIR emission spanning a greater spatial extent. We also overplot the effective radii of several key physical quantities of the galaxy: stellar mass, dust mass and cool, dense gas mass. We define cool, dense gas using the criteria $T<300\,{\rm K}, n_{H} > 10\, \rm{cm}^{-3}$. This has been shown to be a reasonable proxy for molecular gas by \cite{Orr2018}. In this example, and in general, the cool gas mass and the dust mass are more spatially extended than the stellar mass. The sizes of the dust continuum emission tend to be between those of the dust mass and the stellar mass. Shorter-wavelength FIR emission, corresponding to hotter dust, tends to be more compact, whereas longer-wavelength FIR emission better traces the extended cold gas and dust.\\
\indent The 870-\micron\, (345-GHz) observed-frame emission probed by ALMA Band 7 is frequently used to study the dust continuum emission of high-redshift galaxies \citep[e.g.][]{Barro2016,Chen2017,Simpson2017b}. For simplicity, from here on, we consider only the 870-\micron\, flux in our discussion of the spatial extent of the dust continuum emission. However, we note that we consistently find that emission at longer rest-frame FIR wavelengths is more extended, as shown in Figure \ref{fig:A1_79_SED}.

\begin{figure*} 
	\centering
	\includegraphics[scale=1.03]{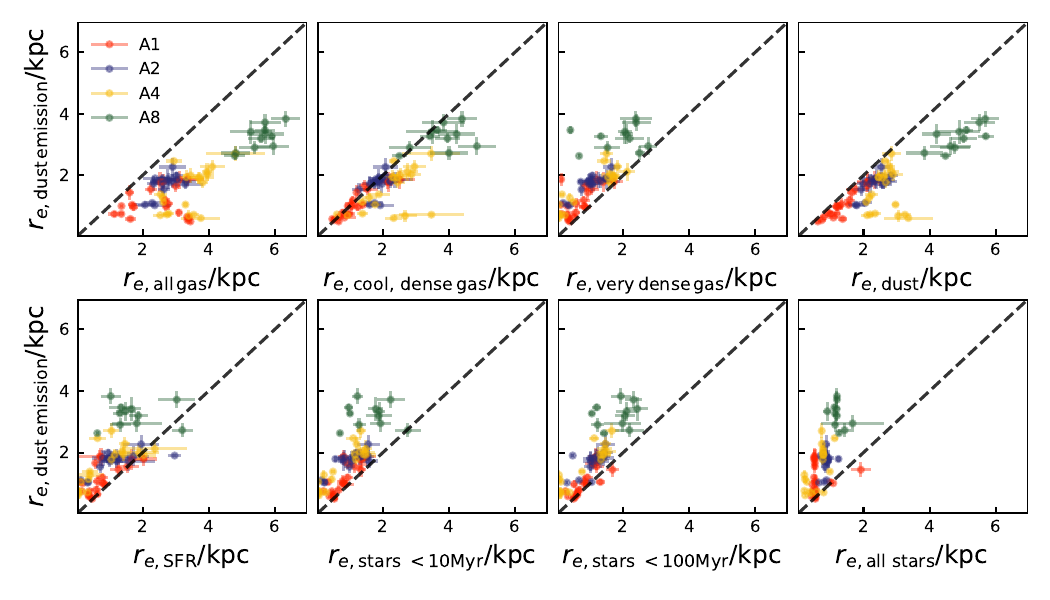}
	\caption{The effective radius of the simulated observed-frame 870-\micron\, dust continuum emission versus those of all gas, cool gas ($T<300\,{\rm K}, n_{H} >10\, \rm{particles}/\rm{cm}^{3}$), dust, very dense gas ($n_{H} >100\,\rm{particles}/\rm{cm}^{3}$), instantaneous star-formation rate, recently formed stars ($\rm{age}<10\rm{Myr}, <\rm{100Myr}$) and all stars, derived directly from the FIRE-2 simulation snapshots within $0.1R_{\rm{vir}}$. Data points are colour-coded according to their haloes. The dashed black lines show a 1-1 relation. The 870-\micron\, emission is more compact than the cool gas and dust but more extended than the very dense gas and stellar components.}
    \label{fig:big_of_alma_band_7_plot}
\end{figure*}

\subsection{Agreement with compact dust continuum observations}\label{sec:dust_cont_gas}
In Figure \ref{fig:cool_gas_dust_observations}, we compare the sizes of \textcolor{black}{both our predicted 870-\micron\, dust emission and the cool, dense gas to the following ALMA observations of galaxies at similar redshifts}: 860-\micron\ emission and $12\rm{CO}(J=4-3)$ line flux for a $\rm{SFR}>1000M_{\odot}/yr$, $M_{*}\sim10^{11}M_{\odot}$ galaxy at $z=4.3$, from \cite{Tadaki2018} (grey star); 870-\micron\, emission and $12\rm{CO}(J=3-2)$ line flux for two massive ($M_{*}\sim10^{11}M_{\odot}$) galaxies at $z=2.5$, from \cite{Tadaki2017} (black stars); the effective radii for stacked ALMA maps of 870-\micron\, and $12\rm{CO}(J=3-2)$ emission for 16 ALESS galaxies at $z=2.5\pm0.2$, from \cite{Rivera2018} (black circle); 870-\micron\, emission and $12\rm{CO}(J=3-2)$ line flux for a  $\rm{SFR}\sim500 M_{\odot}/yr$, $M_{*}\sim2\times10^{11}M_{\odot}$ galaxy at $z=2.12$, from \cite{Chen2017} (black square); FIR emission (from APEX/LABOCA and {\it{Herschel}}) and $12\rm{CO}(J=3-2)$ at $z=6.9$, from \cite{Strandet2017} (black triangle); stacked  870-\micron\, and [CII] emission for $z\sim4.5$ galaxies, from \cite{Cooke2018a} (purple star); $1.1\rm{mm}$ (from NOEMA) and $\rm{H}\alpha$ emission, mapped for a $M_{*}\sim7\times10^{10}M_{\odot}$ at $z=1.25$, from \cite{Nelson2019} (purple circle). Our derived 870-\micron\, effective radii are $\sim 0.5-4\rm{kpc}$. This is in excellent agreement with observations at a range of redshifts \citep[e.g.][]{Ikarashi2015,Iono2016,Simpson2015a,Hodge2016}. In line with the observations, the snapshots of each of the four FIRE-2 galaxies tend to lie below the 1-1 line (dashed black), i.e. the dust continuum emission is more compact than the `molecular' gas. Our predictions agree fairly well with the observational results for the small number of high-redshift galaxies that have been observed in both dust continuum and $\rm{CO}$ at high spatial resolution. However, the few observations that do exist tend to have slightly more compact dust emission at fixed molecular gas effective radius than our predictions. This could be due to the definition of `molecular' gas within the simulation (different temperature and density cuts yield slightly different `molecular' gas sizes, and hence slopes; \textcolor{black}{`very dense gas' is much more compact, as shown in Figure \ref{fig:big_of_alma_band_7_plot}}), or due to the selection of observational samples.

\subsection{How does the dust continuum emission trace the physical properties of galaxies?}\label{sec:analysis}
The dust continuum emission is frequently used as an indicator of both SFR and dust mass. However, until now, the spatial extent of these physical and observable quantities has not been studied consistently with simulated galaxies. Motivated by the clear differences in effective radii of the dust continuum emission, stellar mass, dust mass, and gas mass found for individual FIRE-2 galaxies (see the right-hand panel of Figure \ref{fig:A1_79_SED}), we set out to identify which physical properties are best reflected by the spatial extent of the dust continuum emission. \\
\indent In Figure \ref{fig:big_of_alma_band_7_plot}, we plot the effective radius of the simulated observed-frame 870-\micron\, dust continuum emission versus those of eight different physical quantities derived directly from the FIRE-2 simulations. The 870-\micron\, emission is more compact than the total gas component, but only slightly more compact than the cool, dense gas. The dust emission is also more compact than the dust mass distribution for all FIRE-2 snapshots. We also consider how the spatial extent of dust emission correlates with that of the total stellar mass, recently formed stellar mass (within 10 Myr and 100 Myr) and instantaneous SFR. We find that the emission from dust is more extended than all of these stellar quantities and than the densest gas ($n_{H}>100\,\rm{particles}/\rm{cm}^{3}$). The extent of the dust emission appears to correlate more tightly with the extent of the most recently formed stars than the total stellar mass. This reflects the role of young stars as the primary heating source for the dust grains that reprocess their short-wavelength light, as will be discussed in detail in the next two subsections. Note also that stellar mass appears to assemble inside-out, with the more recently formed stars spanning a larger half-mass radius than the total stellar component.

\subsection{The role of star-formation in determining the extent of the dust continuum emission}\label{sec:pca}
The observed-frame submm emission depends on the effective dust temperature, which is sensitive to the luminosity absorbed by dust, the dust mass, and possibly the geometry of the system \citep[e.g.][]{Misselt_2001,Hayward2011,Safarzadeh2016}.
Consequently, we expect that the spatial extent of the dust continuum will depend on those of the young stars and dust.
To quantify this dependence, we use a Principal Component Analysis (PCA). PCA is a statistical technique used to describe the variance within a dataset. Variables are converted into a set of uncorrelated, orthogonal principal components. The first component reveals the direction of maximum variance, which describes the key correlation between variables within the dataset. Successive components account for less of the variance of the population. Some latter components may be dominated by noise, leaving the data decomposed into fewer dimensions. PCA has been used in a number of recent galaxy evolution studies, including in studies of the mass-metallicity relation \citep{Lagos2015,Bothwell2016,Hashimoto2018} and to study the quenching of galaxies within the EAGLE simulations \citep{Cochrane2018}. \\
\indent Here, we use PCA to study the relationship between the effective radii of the dust component, SFR and dust continuum emission. We use the PCA python tool {\it{scikit.learn}} to perform the PCA analysis. Each variable is normalized to its mean and scaled to unit variance before performing the PCA. For each snapshot studied, we construct a vector of the form [$r_{\rm{eff,\,\,dust\,\,emission}}$, $r_{\rm{eff,\,\,dust}}$, $r_{\rm{eff,\,\,stars\,\,<100Myr}}$] and supply all of these vectors to the PCA.\\
\indent We find that the first principal component, [0.62, 0.56, 0.55], almost entirely describes the dataset, accounting for $83\%$ of the variance of the sample. This primary correlation indicates that all three variables -- the effective radii of dust emission, dust mass and recently formed stars -- are positively correlated. The effective radius of the dust emission exhibits a strong correlation with the effective radius of the dust mass itself and with the effective radius of the recently formed stars. This result confirms our physical intuition that the sub-mm flux density (and thus the effective radius of the dust continuum emission) is sensitive to both the SFR and dust mass \citep{Hayward2011,Safarzadeh2016}. The second principal component, [0.03, 0.69,-0.73], accounts for $15\%$ of the scatter, essentially all of the rest. This component represents the scatter in the relation between the effective radii of the dust mass and the recently formed stars. Note that there is, therefore, little scatter in the strong correlations between the effective radius of the dust emission and the dust mass and between the effective radius of the dust emission and the recently formed stars. 

\begin{figure*} 
\begin{subfigure}{0.5\textwidth}
\includegraphics[scale=0.65]{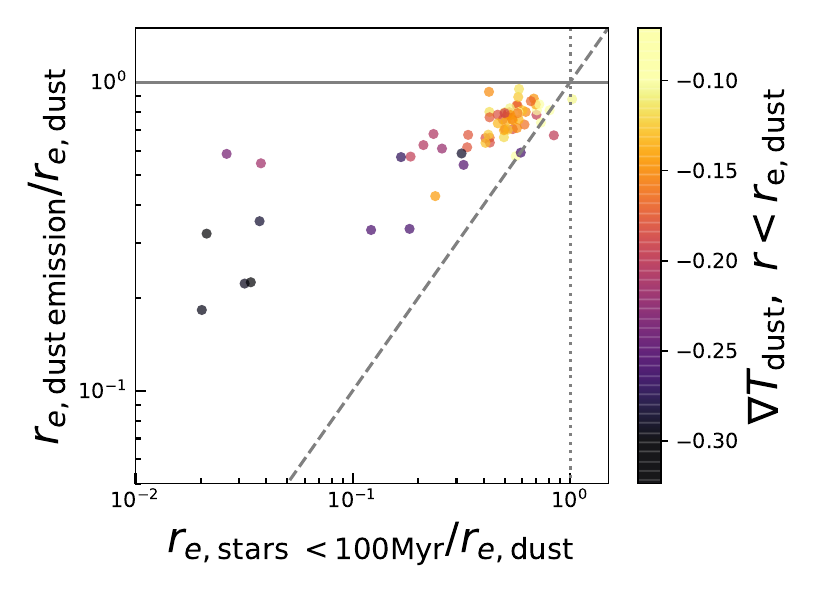}
\caption{} \label{fig:8a}
\end{subfigure}
\hspace*{-0.2cm}
\begin{subfigure}{0.5\textwidth}
\includegraphics[scale=0.6]{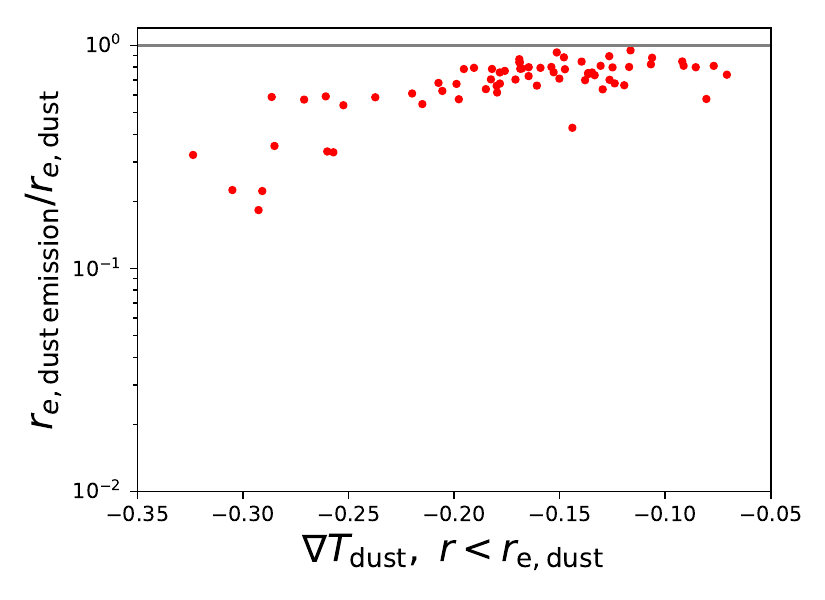}
\caption{} \label{fig:8b}
\end{subfigure}
\begin{subfigure}{0.5\textwidth}
\includegraphics[scale=0.6]{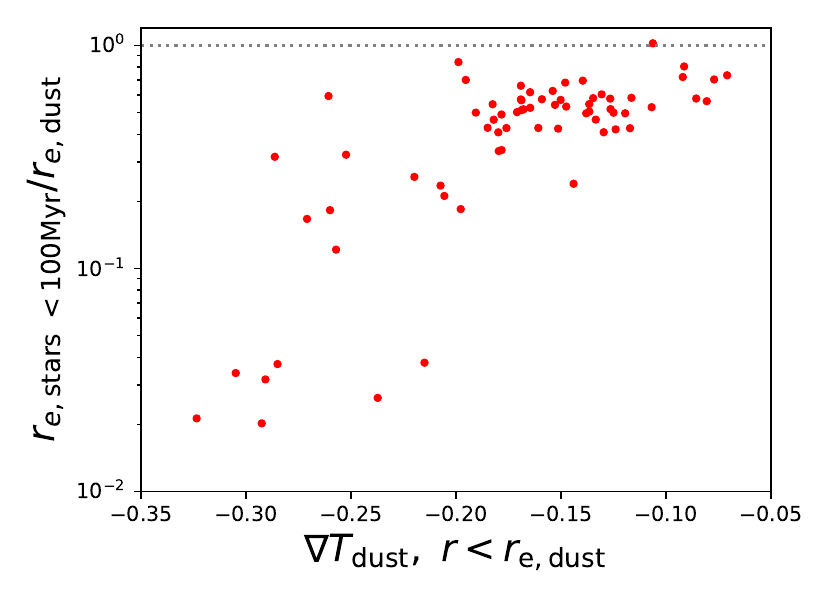}
\caption{} \label{fig:8c}
\end{subfigure}
\hspace*{-0.2cm}
\begin{subfigure}{0.5\textwidth}
\includegraphics[scale=0.6]{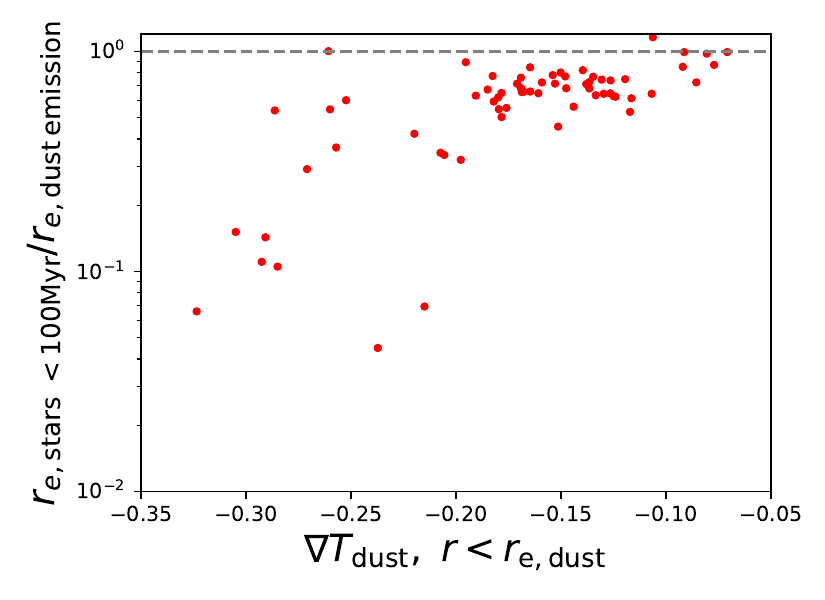}
\caption{} \label{fig:8d}
\end{subfigure}
\caption{The role of dust temperature gradient in determining the size of the observed-frame 870-\micron\, dust continuum emission. (a) The x-axis shows the effective radius of recently formed stars ($\rm{age<100Myr}$), divided by effective radius of the dust mass itself, for each snapshot studied in this work. The y-axis shows the effective radius of the observed-frame dust emission, divided by effective radius of the dust mass. The solid grey line shows $r_{e,\,\rm{dust\,\,emission}}=r_{e,\,\rm{dust}}$, the dashed grey line shows $r_{e,\,\rm{dust\,\,emission}}=r_{e,\,\rm{stars\,\,<100Myr}}$, and the dotted grey line shows $r_{e,\,\rm{dust}}=r_{e,\,\rm{stars\,\,<100Myr}}$. Each snapshot is colour-coded by the dust temperature gradient, calculated within concentric shells on the three-dimensional dust grid. Snapshots where dust continuum emission is particularly compact with respect to the dust mass distribution also display even more compact recent star-formation. This is associated with steeper dust temperature gradients. Panels (b), (c) \& (d) show ratios of the effective radii of dust emission, dust mass and recently formed stars, against  dust temperature gradient. For gradients steeper than $\nabla T_{\rm{dust}}\sim-0.2$, dust continuum emission is a poor tracer of the spatial extent of the dust mass or recently formed stars.}
\label{fig:dust_of_colourbar_T_gradient_cut}
\end{figure*}
\subsection{The physical drivers of compact dust emission: dust temperature gradients}\label{sec:temperature_gradients}

\indent In Section \ref{sec:pca}, we found that the effective radius of the dust continuum emission is correlated with both the effective radius of the dust mass itself, and the effective radius of the recently formed stars. In this final section, we investigate the physical drivers of these correlations, in particular the role of dust heating. In Figure \ref{fig:dust_of_colourbar_T_gradient_cut}, we plot the effective radius of the 870-\micron\, dust continuum emission against that of the recently formed stars, normalising both quantities by the effective radius of the dust mass. If the dust continuum emission traced the dust mass directly, all points would lie at $r_{e,\rm{dust\,emission}}/r_{e,\rm{dust}}=1$. While $r_{e,\rm{dust\,emission}}/r_{e,\rm{dust}}$ is close to 1 for most snapshots, we find a clear slope in this relation: galaxies with low $r_{e,\rm{dust\,emission}}/r_{e,\rm{dust}}$ (which have particularly compact dust emission with respect to the total dust mass distribution) also have recent star-formation which is much more compact than the dust. This reflects the dual role of the dust mass and star-formation in determining the spatial extent of the dust continuum emission.\\
\indent We have identified that the spatial scales of dust and recently formed stars play roles in determining the extent of the dust continuum emission, with very compact dust emission appearing to correlate with compact recent star-formation. Now, we use the dust mass-weighted temperatures of the grid cells output by {\sc skirt} to explore the physical drivers of particularly compact dust emission. \textcolor{black}{The temperature of the dust in our simulated galaxies varies significantly as a function of distance from the galaxy's centre. Typical dust temperatures are $\sim20-40\rm{K}$ outside the central kpc, and $\sim40-70\rm{K}$ inside the central kpc. Here, we study the relationship between dust temperature gradients and the spatial extent of dust emission.}
For each snapshot, we construct thin concentric shells about the halo centre, out to the effective radius calculated for the dust mass.
For each shell, we calculate the mean dust temperature, $T_{\rm{dust}}$, and mean dust particle position, $r_{\rm{dust}}$. We then fit the following relation: $\log_{10}T_{\rm{dust}} = \nabla T_{\rm{dust}} \log_{10}r_{\rm{dust}}+C$, where the units of $T_{\rm{dust}}$ are K and the units of $r_{\rm{dust}}$ are kpc. This yields a dust temperature gradient (strictly a power-law index), $\nabla T_{\rm{dust}}$, for each galaxy snapshot. The snapshots in Figure \ref{fig:dust_of_colourbar_T_gradient_cut}(a) are colour-coded by this gradient. It is clear that those galaxies with compact dust emission and recent star-formation also have the steepest dust temperature gradients. The other panels of Figure \ref{fig:dust_of_colourbar_T_gradient_cut} show ratios of the effective radii of the dust mass, dust continuum emission and recently formed stars. For gradients flatter than $\nabla T_{\rm{dust}}\sim-0.2$, the dust continuum emission broadly traces the extent of both recently formed stars and dust mass. For steeper gradients, below $\nabla T_{\rm{dust}}\sim-0.2$, the recently formed stars are substantially more compact than the dust mass. This causes the ratio of absorbed luminosity to dust mass to be higher in the galaxy centre than at larger radii, leading to the steep dust temperature gradients. At these low values of $\nabla T_{\rm{dust}}$, the spatial extent of the dust continuum emission is not a good approximation of the distribution of dust mass or recently formed stars.\\
\indent These results confirm that the spatial extent of the dust continuum emission is sensitive to the scales of recent star-formation because of the effects of dust heating. Thus, inferring spatial scales of star-formation or dust mass from observations of dust continuum emission is non-trivial. This may be done more robustly using measurements of dust temperature gradients within galaxies. Studies of local galaxies show that dust temperature distributions can be derived from spatially-resolved imaging in multiple bands \citep[e.g.][]{Galametz2012}.
\section{Conclusions}\label{sec:conclusions}
We have performed radiative transfer modelling on a subset of rest-frame FIR-bright redshift snapshots of four massive galaxies drawn from the FIRE-2 simulations presented in \citet{Angles-Alcazar2017}. 
% DAA: adding AA+2017 for quick reference to the simulation paper
The simulated galaxies have stellar masses $7\times10^{10}-3\times10^{11}M_{\odot}$, and reside in dark matter haloes of mass $M_{\rm{halo}}\sim10^{12.5} M_{\odot}$ at $z=2$. Our modelling yields full rest-frame FUV-FIR SEDs as well as maps of the emission in $\sim 100$ wavebands, resolved to $25\rm{pc}$ scales. We find clear differences between the morphologies of the same galaxies in the different wavebands, with shorter-wavelength emission (UV/optical) frequently appearing clumpy and extended. We find that this is due to the emitted short-wavelength light tracing `holes' in the dust distribution. At rest-frame FIR wavelengths, the galaxies tend to have more regular morphologies. \\
\indent The primary focus of our study is the spatial extent of the dust continuum emission. For the snapshots we study, the FIRE-2 galaxies have observed-frame 870-\micron\, fluxes of up to $\sim2\,\rm{mJy}$ at $z\sim1-5$. These simulated FIR-bright galaxies exhibit very compact dust continuum emission, with effective radii of $\sim0.5-4\rm{kpc}$, in line with existing observations of star-forming galaxies at these redshifts. At rest-frame FIR wavelengths (which we can probe with ALMA for high-redshift galaxies), longer-wavelength emission tends to be more extended because the dust tends to be hotter in the central regions of galaxies, and the shorter-wavelength emission is dominated by hotter grains.\\
\indent We also compare the spatial extent of the  870-\micron\, dust continuum emission to that of several key intrinsic physical quantities, including the dust, gas and stellar components. In both simulated and observed galaxies, the dust continuum emission is more compact than the cool, dense gas and the dust, but more extended than the stellar component. Extracting only recently formed ($\rm{age}<100\rm{Myr}$) stars from the simulations enables us to study the role of recent star-formation in determining the spatial extent of the dust emission. We find that in some snapshots, the simulated galaxies exhibit extremely compact dust emission ($\sim0.5\rm{kpc}$). This emission appears to be driven by particularly compact recent star-formation. Overall, the spatial extent of the dust continuum emission correlates with those of both the dust mass and the recently formed stars. Physically, this makes sense: the dust that emits strongly in the FIR is predominantly heated by young massive stars (as these simulations do not contain AGN), and when the recent star-formation is particularly compact, the central regions of the galaxy have steeper dust temperature gradients and consequently more compact emission. In such systems, constraints on the dust temperature gradient are necessary to infer the spatial extents of the young stars and dust reliably. This work thus motivates multiband ALMA observations to constrain the dust temperature gradients of observed galaxies.

\section*{Acknowledgements}
This work was initiated as a project for the Kavli Summer Program in Astrophysics held at the Center for Computational Astrophysics of the Flatiron Institute in 2018. The program was co-funded by the Kavli Foundation and the Simons Foundation. We thank them for their generous support. RKC acknowledges funding from STFC via a studentship and thanks Philip Best for helpful discussions and comments. DAA acknowledges support from a Flatiron Fellowship. The Flatiron Institute is supported by the Simons Foundation. DK was supported by NSF grant AST-1715101 and the Cottrell Scholar Award from the Research Corporation for Science Advancement. RF acknowledges financial support from the Swiss National Science Foundation (grant number 157591). CAFG was supported by NSF through grants AST-1517491, AST-1715216, and CAREER award AST-1652522; by NASA through grant 17-ATP17-0067; by CXO through grant TM7-18007; by STScI through grant HST-AR-14562.001; and by a Cottrell Scholar Award from the Research Corporation for Science Advancement. The simulations were run using XSEDE, supported by NSF grant ACI-1053575, and Northwestern University's compute cluster `Quest'.

\bibliographystyle{mnras}
\bibliography{Edinburgh_ALMA2}

%\bsp
\label{lastpage}
\end{document}